# No or diffuse phase transition with temperature in one-dimensional Ising model?

Yi-Neng Huang[1,2]*, Li-Li Zhang[1]

[1] Xinjiang Laboratory of Phase Transitions and Microstructures in Condensed Matters, College of Physical Science and Technology, Yili Normal University; Yining, China.

[2] National Laboratory of Solid State Microstructures, School of Physics, Nanjing University; Nanjing, China.

*Corresponding author. Email: ynhuang@nju.edu.cn

**Abstract:** For nearly a century since Ising model was proposed in 1925, it is agreed that there is no phase transition with temperature in the one-dimensional based on no global spontaneous magnetization in whole temperature region. In this paper, the exact calculation of local spontaneous magnetization shows that a diffuse phase transition with temperature occurs in one-dimensional Ising model. In addition, although diffuse phase transition phenomenon is common in the systems of heterogeneous-components and grains etc., there is no accurate prediction of corresponding theoretical models so far, so the present works lay the theoretical foundation of this kind of phase transition.

**Main Text:** In the nearly 100 years since Ising model (IM) was proposed in 1925 (*1*), it is agreed that there is no temperature dependent phase transition in the one-dimensional. This is because the global spontaneous magnetization of the model system is zero in whole temperature range, that is, there is no global spontaneous magnetization (*2*) as shown in the supplementary material (SM).

However, the absence of global spontaneous magnetization does not deny the existence of short-range local spontaneous magnetization in one-dimensional-IM (1D-IM). In this paper, the local spontaneous magnetization with temperature and size in the model is calculated accurately, and the results show that 1D-IM has a diffuse phase transition with temperature (*3-7*).

The Hamiltonian ($H_{1D-IM}$) of 1D-IM is,

$$H_{1D-IM} = \lim_{N \to \infty} \left[ -J \sum_{i=1}^{N-1} \sigma_i \sigma_{i+1} \right] \quad (1)$$

in which $\sigma_i$ is the ith spin and $\sigma_i = \pm 1$, $J$ the interaction energy constant between the nearest-neighbor spins, and $N$ the total number of spins in the model system (fig. S1).

In the model, the local magnetic moment ($s_l^r$) including $l$ nearest-neighbor spins is,

$$s_l^r \equiv \mu \sum_{i=0}^{l-1} \sigma_{r+i} \quad (2)$$

where $\mu$ is the magnetic moment of a spin, and $r$ expresses an arbitrary reference site.

To describe the temperature dependence of the amplitude of $s_l^r$ (excluding the orientations corresponding to its signs) in this paper, the local spontaneous magnetization ($m_s^l$) is defined as (SM),

$$m_s^l \equiv \frac{1}{l}\sqrt{\lim_{n\to\infty}\frac{1}{Z_n}\sum_{\sigma_i=\pm 1,\cdots\sigma_n=\pm 1}(s_l^r)^2 \exp\left[\frac{J}{k_BT}\sum_{i=1}^{n-1}\sigma_i\sigma_{i+1}\right]}$$

$$= \frac{\mu}{l}\sqrt{2\left[\frac{l-\gamma^l}{1-\gamma}-\frac{\gamma(1-\gamma^{l-1})}{(1-\gamma)^2}\right]-l} \quad (3)$$

here $k_B$ is Boltzmann constant, $T$ the temperature of the heat bath in which the one-dimensional spin chain is located, $Z_n$ the partition function of the spin orientation ensemble of 1D-IM (SM), and $n$ the spin number of the subsystems in the ensemble (fig. S1), and $\gamma \equiv \tanh\left(\frac{J}{k_BT}\right)$.

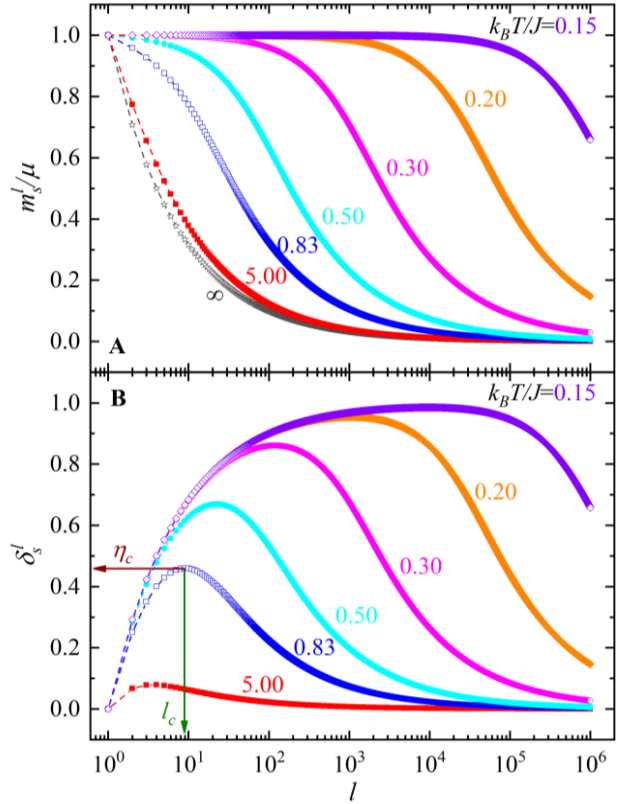

**Fig. 1. For series $T$, (A) local spontaneous magnetization ($m_s^l$) and (B) relative local spontaneous magnetization ($\delta_s^l$) vs local spin number ($l$) in one-dimensional Ising model.**

Figure 1A shows $m_s^l$ vs $l$ for series $T$, and it can be seen that: (i) At high temperature (e.g. $T = 5.00 J/k_B$), $m_s^l$ decreases rapidly with increasing $l$, which indicates that the spatial scale of local spontaneous magnetization is small; and (ii) At low temperature (e.g. $T = 0.15 J/k_B$), $m_s^l \to \mu$ in a large range of $l$, which states clearly that the local spontaneous magnetization regions not only has a large spatial scale, but also almost all the spins in the regions have the same orientation.

According to Landau theory (8), the order parameter is the essential to phase transition, which characterizes the relative change of the low to high temperature phase. Therefore, in order to describe the relative variation of $m_s^l$ to $m_s^l(T \to \infty)$, the relative local spontaneous magnetization ($\delta_s^l$) is introduced here,

$$\delta_s^l \equiv \frac{m_s^l - m_s^l(T \to \infty)}{\mu} \tag{4}$$

At series $T$, $\delta_s^l$ vs $l$ is shown in Figure 1B, which indicates that for all temperatures, $\delta_s^l$ has a single diffuse peak as a function of $l$. In this paper, the maximum value of $\delta_s^l$ is expressed as $\eta_c$, and the corresponding value of $l$ as $l_c$. Obviously, $\eta_c$ and $l_c$ can be used as the characteristic parameters to describe the local spontaneous magnetization and its spatial size, so here $\eta_c$ is called the characteristic spontaneous magnetization and $l_c$ the characteristic spatial size of $\eta_c$ in 1D-IM. Moreover, the dispersion of the $\delta_s^l$ peak shows that both the size of the local spontaneous magnetization regions and its internal magnetization have obvious distributions.

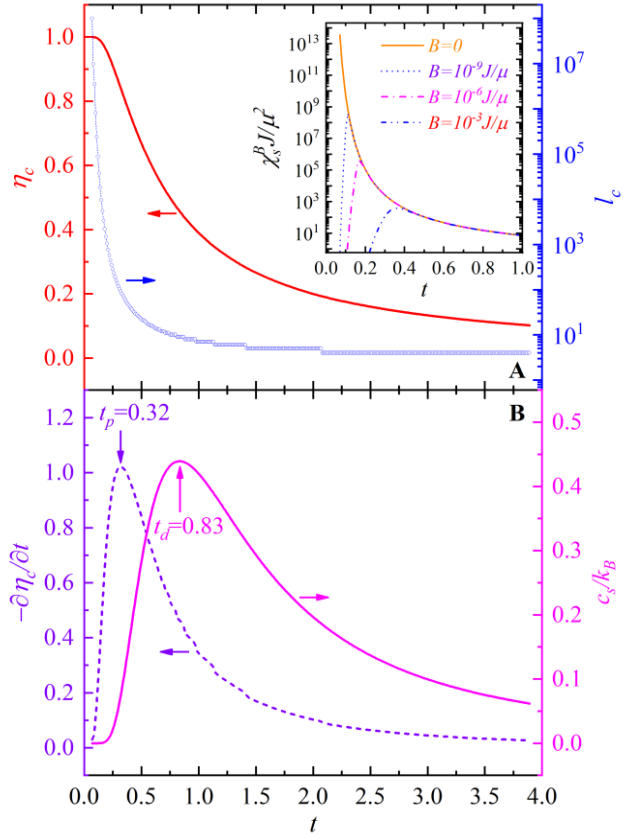

**Fig. 2.** (A) characteristic spontaneous magnetization ($\eta_c$) and its characteristic spatial size ($l_c$), as well as (B) $-\partial \eta_c/\partial t$ and specific heat per spin ($c_s$) in 1D-IM vs reduced temperature ($t \equiv k_B T/J$). Inset of fig. 2A shows static susceptibility ($\chi_s^B$) per spin vs $t$ for a series of external magnetic fields ($B$).

$\eta_c$ and $l_c$ vs $T$ (fig. 2A) show that with decreasing $T$: (i) $\eta_c$ first increases slowly, then rapidly, and slowly again. It should be noted that at high temperature, $\eta_c$ is still not zero (e.g. $\eta_c = 0.10$ when $T = 3.82 J/k_B$). In addition, $\eta_c \to 1$ at nonzero low temperature (e.g. $\eta_c = 0.9999$ for $T = 0.07 J/k_B$); and (ii) $l_c$ first increases slowly and then rapidly. For example, $l_c = 4$ when $T = 3.82 J/k_B$, and $l_c = 10^8$ for $T = 0.07 J/k_B$ (if the lattice parameter of 1D-IM is assumed to be 0.5 nanometers, the characteristic size of the local spontaneous magnetization region will reach the macroscopic ~ 5 centimeters). These results state clearly that in 1D-IM, there exists a diffuse transition between the nanoscale regions of small spontaneous magnetization at high temperature and the macroscopic domains of large spontaneous magnetization at low temperature.

By comparing the order parameters (*4-7*), specific heat (*9-11*), and domain structure evolution (*12-14*) of existing diffuse phase transition, with the diffuse variation of $\eta_c$, the diffuse peak (*1, 2*) of specific heat ($c_s$) per spin (SM and fig. 2B), and the diffuse transition between the nanoscale regions of local spontaneous magnetization to the macroscopic domains as a function of $T$, it can be concluded that a diffuse phase transition with temperature occurs in 1D-IM.

According to the method of reference (*15*), the temperature corresponding to the maximum of $-\frac{\partial \eta_c}{\partial T}$ is defined as the characteristic temperature ($T_p$) of diffuse phase transition in 1D-IM, and it is obtained $T_p = 0.32 J/k_B$ (fig. 2B). It should be pointed out that $T_p$ is lower than the peak temperature ($T_d = 0.83 J/k_B$) of $c_s$. For second-order or continuous phase transition, the peak temperatures of specific heat and the negative of the differential of order parameter to $T$ are equal to each other (*8, 16, 17*), so the authors thinks that the difference between $T_p$ and $T_d$ just reflects the dispersion of the diffuse phase transition. In order to describe this dispersion, the dispersion degree ($\varphi$) of the diffuse phase transition is proposed as,

$$\varphi \equiv \frac{T_d - T_p}{T_d} = 0.61 \tag{5}$$

It is worth noting that the static susceptibility ($\chi_s$) per spin in 1D-IM (inset of fig. 2A and SM) always increases rapidly with decreasing $T$ (*1, 2*), instead of the $\lambda$-type peak of second-order phase transition (*8*), which is one of the key evidences that no phase transition with temperature exist in this model.

In order to further explore the micro mechanism of the above characteristic of $\chi_s$, the static susceptibility ($\chi_s^B$) per spin in 1D-IM as a function of $T$ in a fixed external magnetic field ($B$) is calculated (SM), as shown in the inset of Figure 2A. We can see that for finite small $B$, there is a single diffuse peak of $\chi_s^B$ with $T$, and the peak temperature moves to high temperature with the increase of $B$. This is due to that very small $B$ can saturate the magnetization of 1D-IM at low temperature (as shown in fig. S2, the saturation $B \sim 10^{-9} J/\mu$ for $T = 0.10 J/k_B$), while the saturated magnetization leads to a smaller value $\chi_s^B$. Because the saturation magnetization corresponds to the single domain state of the model, the increase of $\chi_s$ at low temperature is caused by the movement of domain walls (*18, 19*).

In particular, because the measurement magnetic field used in experiments is always finite, the susceptibility peak (inset of fig. 2A) will appear in 1D-IM as long as the experimental measurement is carried out. In other words, the continuously increasing characteristic of the theoretically predicted $\chi_s$ ($\chi_s^{B=0}$) with decreasing $T$ cannot be measured directly, which is only ideal value.

Although diffuse phase transition is common in component-heterogeneous (*4-7, 20, 21*) and granular systems (*22, 23*), there is no accurate calculation of the corresponding theoretical model so far (*15*). Therefore, the exact results in this paper lay the theoretical foundation of this kind of phase transition.

Weiss mean field theory (*24*) is always thought to be not suitable for 1D-IM because it predicts that there is a phase transition with $T$ in this system. The present and relevant (*2, 16, 17*) results show that this theory is valid for IM of all dimensions in judging whether there exists phase transition, although its predicting phase transition behaviors, such as the transition temperature and the subtle characteristics of order parameter and specific heat, are different from the exact solutions (fig.2 and *2, 16, 17*). It also shows that Weiss-type mean field approximation (*15*) is an effective and feasible method for phase transition models which are too complex and difficult to get their exact solutions.

# Supplementary Materials for

## No or diffuse phase transition with temperature in one-dimensional Ising model?

Yi-Neng Huang[1,2]*, Li-Li Zhang[1]

Correspondence to: ynhuang@nju.edu.cn

Calculation of global spontaneous magnetization in 1D-IM
At present, the method to calculate the global spontaneous magnetization of Ising model (2, 17) is as follows: (i) Applying a static external field ($B$) to the model system; (ii) Constructing the spin orientation ensemble corresponding to the model; (iii) Calculating the global magnetization of the ensemble based on Boltzmann principle; (iv) Let $B \to 0$, and the obtained magnetization is the global spontaneous magnetization of the system.

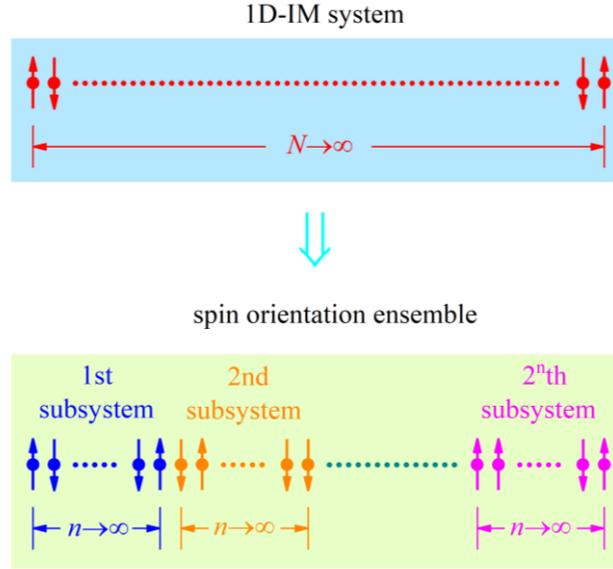

**Fig. S1. Diagrammatic sketch of ID-IM system and construction of corresponding spin orientation ensemble.**

According to this method, the Hamiltonian ($H_{1D-IM}^B$) of 1D-IM with $B$ is,

$$H_{1D-IM}^B = \lim_{N \to \infty} \left[ -J \sum_{i=1}^{N-1} \sigma_i \sigma_{i+1} - \mu B \sum_{i=1}^{N} \sigma_i \right] \tag{S1}$$

where $\sigma_i$ is the ith spin and $\sigma_i = \pm 1$, $J$ the interaction energy constant between nearest-neighbor spins, $\mu$ the magnetic moment of a spin, and $N$ the number of spins in the model system.

The spin orientation ensemble corresponding to 1D-IM is constructed as shown in Figure S1, i.e. the whole chain of $N \to \infty$ spins is divided into $2^n$ sub-chains (subsystems) of $n \to \infty$ spins with different spin orientation configurations.

The Hamiltonian ($H_n^B$) of the subsystem with $B$ (the endpoint effect of the subsystems can be ignored for $n \to \infty$) is,

$$H_n^B = -J \sum_{i=1}^{n-1} \sigma_i \sigma_{i+1} - \mu B \sum_{i=1}^{n} \sigma_i \tag{S2}$$

and based on Boltzmann principle, the partition function ($Z_n^B$) of the ensemble is,

$$Z_n^B \equiv \sum_{\sigma_i=\pm 1,\cdots\sigma_n=\pm 1} \exp\left[v \sum_{i=1}^{n-1} \sigma_i\sigma_{i+1} + w \sum_{i=1}^{n} \sigma_i\right] \tag{S3}$$

here $v \equiv \frac{J}{k_B T}$, $w \equiv \frac{\mu B}{k_B T}$, $T$ is temperature, and $k_B$ Boltzmann constant.

From eq. S3, the global magnetization ($m_B$) per spin in 1D-IM is,

$$m_B = \lim_{n\to\infty}\left[\frac{\mu}{nZ_n}\frac{\partial Z_n}{\partial w}\right] \tag{S4}$$

According to $\exp(v\sigma_i\sigma_{i+1}) = \cosh(v)(1+\gamma\sigma_i\sigma_{i+1})$ and $\exp(w\sigma_i) = \cosh(w)(1+\alpha\sigma_i)$, we get,

$$Z_n^B = \cosh^{n-1}(v)\cosh^n(w)Q_n^B \tag{S5}$$

in which $\gamma \equiv \tanh(v)$, $\alpha \equiv \tanh(w)$, and,

$$Q_n^B \equiv \sum_{\sigma_i=\pm 1,\cdots\sigma_n=\pm 1} \prod_{i=1}^{n-1}(1+\gamma\sigma_i\sigma_{i+1}) \prod_{i=1}^{n}(1+\alpha\sigma_i) \tag{S6}$$

Defining,

$$Y_n^B \equiv \sum_{\sigma_i=\pm 1,\cdots\sigma_n=\pm 1} \prod_{i=1}^{n-1}(1+\gamma\sigma_i\sigma_{i+1}) \prod_{j=1}^{n}(1+\alpha\sigma_i)\sigma_n \tag{S7}$$

Obviously, $Q_1^B = 2$ and $Y_1^B = 2\alpha$.

From $\sum_{\sigma_n=\pm 1}(1+\gamma\sigma_{n-1}\sigma_n)(1+\alpha\sigma_n) = 2(1+\alpha\gamma\sigma_{n-1})$ and $\sum_{\sigma_n=\pm 1}(1+\gamma\sigma_{n-1}\sigma_n)(1+\alpha\sigma_n)\sigma_n = 2(\alpha+\gamma\sigma_{n-1})$, we obtain,

$$\begin{bmatrix}Q_n^B \\ Y_n^B\end{bmatrix} = 2\begin{bmatrix}1 & \alpha\gamma \\ \alpha & \gamma\end{bmatrix}\begin{bmatrix}Q_{n-1}^B \\ Y_{n-1}^B\end{bmatrix} \tag{S8}$$

and,

$$\begin{bmatrix}Q_n^B \\ Y_n^B\end{bmatrix} = 2^{n-1}\begin{bmatrix}1 & \alpha\gamma \\ \alpha & \gamma\end{bmatrix}^{n-1}\begin{bmatrix}Q_1^B \\ Y_1^B\end{bmatrix} = 2^n\begin{bmatrix}1 & \alpha\gamma \\ \alpha & \gamma\end{bmatrix}^{n-1}\begin{bmatrix}1 \\ \alpha\end{bmatrix} \tag{S9}$$

The eigenvalues of the square matrix $\begin{bmatrix}1 & \alpha\gamma \\ \alpha & \gamma\end{bmatrix}$ is determined by the following characteristic equation,

$$\begin{vmatrix}1-\lambda_B & \alpha\gamma \\ \alpha & \gamma-\lambda_B\end{vmatrix} = 0 \tag{S10}$$

resulting in,

$$\lambda_B = \frac{1+\gamma \pm \sqrt{(1-\gamma)^2 + 4\gamma\alpha^2}}{2} \tag{S11}$$

i.e. the two values ($\lambda_B^1$ and $\lambda_B^2$) of $\lambda_B$ are, respectively,

$$\lambda_B^1 = \frac{1+\gamma + \sqrt{(1-\gamma)^2 + 4\gamma\alpha^2}}{2}$$
$$\lambda_B^2 = \frac{1+\gamma - \sqrt{(1-\gamma)^2 + 4\gamma\alpha^2}}{2} \tag{S12}$$

From eq. S9 and S12, we get,

$$\begin{bmatrix}Q_n^B \\ Y_n^B\end{bmatrix} = 2^n\begin{bmatrix}U_{11} & U_{12} \\ U_{21} & U_{22}\end{bmatrix}\begin{bmatrix}(\lambda_B^1)^{n-1} & 0 \\ 0 & (\lambda_B^2)^{n-1}\end{bmatrix}\begin{bmatrix}V_{11} & V_{12} \\ V_{21} & V_{22}\end{bmatrix}\begin{bmatrix}1 \\ \alpha\end{bmatrix} \tag{S13}$$

where $\begin{bmatrix}V_{11} \\ V_{21}\end{bmatrix}$ and $\begin{bmatrix}V_{12} \\ V_{22}\end{bmatrix}$ are the eigenvectors corresponding to $\lambda_B^1$ and $\lambda_B^2$, and $\begin{bmatrix}U_{11} & U_{12} \\ U_{21} & U_{22}\end{bmatrix} \equiv \begin{bmatrix}V_{11} & V_{12} \\ V_{21} & V_{22}\end{bmatrix}^{-1}$.

Therefore,

$$\lim_{n\to\infty}\begin{bmatrix}Q_n^B\\Y_n^B\end{bmatrix} = 2^n(\lambda_B^1)^n \begin{bmatrix}U_{11} & U_{12}\\U_{21} & U_{22}\end{bmatrix}\begin{bmatrix}1 & 0\\0 & 0\end{bmatrix}\begin{bmatrix}V_{11} & V_{12}\\V_{21} & V_{22}\end{bmatrix}\begin{bmatrix}1\\\alpha\end{bmatrix}$$
$$= 2^n(\lambda_B^1)^n \begin{bmatrix}V_{11}U_{11} + V_{12}U_{11}\alpha\\V_{11}U_{21} + V_{12}U_{21}\alpha\end{bmatrix} \quad (S14)$$

Based on eq. S4, S12 and S14, we obtain,

$$m_B = \frac{\mu\sinh(w)}{[\sinh^2(w) + \exp(-4v)]^{1/2}} \quad (S15)$$

According to this equation, it is easy to find $m_{B\to 0} \to 0$ for nonzero temperature. $m_B$ vs $B$ at series $T$ is shown in fig. S2.

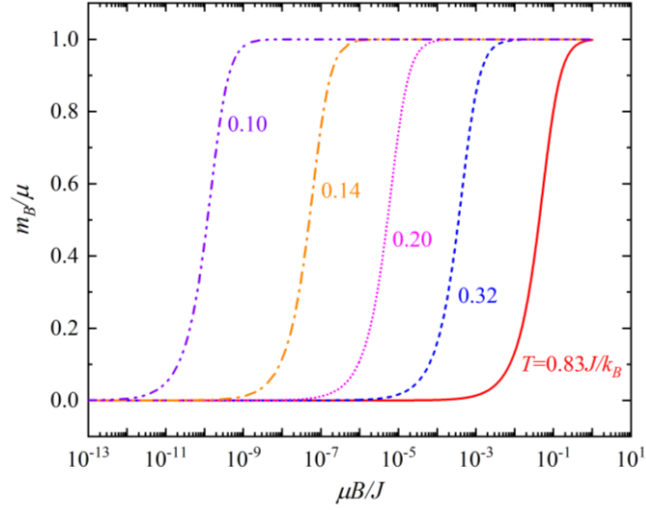

**Fig. S2. Global Magnetization ($m_B$) per spin in 1D-IM vs external magnetic field ($B$) at series temperature ($T$).**

When $B$ is present, the susceptibility ($\chi_s^B$) per spin in 1D-IM is,

$$\chi_s^B \equiv \frac{\partial m_B}{\partial B} \quad (S16)$$

From eq. S15 and S16, we get,

$$\chi_s^B = \frac{\mu^2}{J}\frac{1}{t}\frac{\cosh\left(\frac{b}{t}\right)\exp\left(-\frac{4}{t}\right)}{\left[\sinh^2\left(\frac{b}{t}\right) + \exp\left(-\frac{4}{t}\right)\right]^{3/2}} \quad (S17)$$

here $t \equiv \frac{k_B T}{J}$ is reduced temperature, and $b \equiv \frac{\mu B}{J}$ reduced magnetic field.

Calculation of local spontaneous magnetization in 1D-IM

The Hamiltonian ($H_n$) of the subsystem (fig. S1) without external magnetic field is,

$$H_n = -J\sum_{i=1}^{n-1}\sigma_i\sigma_{i+1} \quad (S18)$$

and the partition function ($Z_n$) of the spin orientation ensemble (fig. S1) is,

$$Z_n \equiv \sum_{\sigma_i=\pm 1,\cdots\sigma_n=\pm 1}\exp\left[v\sum_{i=1}^{n-1}\sigma_i^1\sigma_{i+1}^1\right] = \cosh^{n-1}(v)Q_n \quad (S19)$$

In which,

$$Q_n \equiv \sum_{\sigma_i=\pm 1,\cdots\sigma_n=\pm 1} \prod_{i=1}^{n-1}(1+\gamma\sigma_i\sigma_{i+1}) = 2^n \tag{S20}$$

Let,

$$X_l \equiv \lim_{n\to\infty}\frac{1}{Z_n}\sum_{\sigma_i=\pm 1,\cdots\sigma_n=\pm 1}(s_l^r)^2\exp\left[v\sum_{i=1}^{n-1}\sigma_i\sigma_{i+1}\right] \tag{S21}$$

and according to,

$$(s_l^r)^2 = \mu^2\left\{l + 2\left[\sum_{i=1}^{l-1}\sigma_{r+i-1}^1\sigma_{r+i}^1 + \sum_{i=2}^{l-1}\sigma_{r+i-2}^1\sigma_{r+i}^1 + \cdots \sum_{i=l-1}^{l-1}\sigma_r^1\sigma_{r+l-1}^1\right]\right\} \tag{S22}$$

we obtain,

$$X_l = \mu^2\left[l + 2\sum_{k=1}^{l-1}(l-k)\zeta_k\right] \tag{S23}$$

where $\zeta_k$ is the correlation function between $\sigma_r$ and $\sigma_{r+k}$, i.e.,

$$\zeta_k \equiv \lim_{n\to\infty}\frac{1}{Z_n}\sum_{\sigma_i=\pm 1,\cdots\sigma_n=\pm 1}\sigma_r\sigma_{r+k}\exp\left[v\sum_{i=1}^{n-1}\sigma_i\sigma_{i+1}\right]$$

$$= \lim_{n\to\infty}\frac{1}{Q_n}\sum_{\sigma_i=\pm 1,\cdots\sigma_n=\pm 1}\prod_{i=1}^{n-1}(1+\gamma\sigma_i\sigma_{i+1})\,\sigma_r\sigma_{r+k} \tag{S24}$$

Based on,

$$I_0 \equiv \sum_{\sigma_r=\pm 1}(1+\gamma\sigma_{r-1}\sigma_r)(1+\gamma\sigma_r\sigma_{r+1})\,\sigma_r$$
$$= 2\gamma(\sigma_{r-1}+\sigma_{r+1}) \tag{S25}$$

$$I_1 \equiv \sum_{\sigma_{r+1}=\pm 1}I_1(1+\gamma\sigma_{r+1}\sigma_{r+2})$$
$$= 2^2\gamma(\sigma_{r-1}+\gamma\sigma_{r+2}) \tag{S26}$$

$$\cdots$$

$$I_k \equiv \sum_{\sigma_{r+k}=\pm 1}I_{k-1}(1+\gamma\sigma_{r+k}\sigma_{r+k+1})\sigma_{r+k}$$
$$= 2^{k+1}\gamma(\gamma^{k-1}+\gamma\sigma_{r-1}\sigma_{r+k+1}) \tag{S27}$$

we get,

$$\zeta_k = \lim_{n\to\infty}\frac{1}{Q_n}\sum_{\substack{\sigma_i=\pm 1,\cdots\sigma_{r-1}=\pm 1\\ \sigma_{r+k+1}=\pm 1,\cdots\sigma_n=\pm 1}}I_k\prod_{i=1}^{r-2}(1+\gamma\sigma_i\sigma_{i+1})\prod_{i=r+k+1}^{n-1}(1+\gamma\sigma_i\sigma_{i+1})$$
$$= \gamma^k \tag{S28}$$

From eq. S23 and S28, we obtain,

$$X_l = \mu^2\left[l + 2\sum_{k=1}^{l-1}(l-k)\gamma^k\right] \tag{S29}$$

here $\sum_{k=1}^{l-1}(l-k)\gamma^k$ is the well-known arithmetic-geometric series, and,

$$X_l = \mu^2\left\{2\left[\frac{l-\gamma^l}{1-\gamma}-\frac{\gamma(1-\gamma^{l-1})}{(1-\gamma)^2}\right]-l\right\} \tag{S30}$$

Therefore,

$$m_s^l = \frac{\mu}{l}\sqrt{2\left[\frac{l-\gamma^l}{1-\gamma} - \frac{\gamma(1-\gamma^{l-1})}{(1-\gamma)^2}\right] - l} \tag{S31}$$

and obviously,

$$m_s^l(T \to \infty) = \frac{\mu}{l^{1/2}} \tag{S32}$$

From eq. S18 and S19, the average internal energy ($u$) per spin in 1D-IM is,

$$u = \lim_{n \to \infty}\left[-\frac{J}{nZ_n}\frac{\partial Z_n}{\partial v}\right] = -J\gamma \tag{S33}$$

and the average specific heat per spin ($c_s$) is,

$$c_s = \frac{\partial u}{\partial T} = \frac{k_B}{t^2}(1-\gamma^2) \tag{S34}$$

10